# A study of risk management in cloud computing bank

Dou El Kefel Mansouri[1], Mohamed Benyettou[2]

[1] Department of computer science, University of sciences and technologies Mohamed Boudiaf, USTO
Oran, Usto, Algeria
*mansouri_douelkefel@yahoo.com*

[2] Department of computer science, University of sciences and technologies Mohamed Boudiaf, USTO
Oran, Usto, Algeria
*Med_benyettou@yahoo.fr*

**Abstract**
Cloud computing apparently helps in reducing costs and providing the scheduling optimal level. In practice however it may confront the problem of unavailability of resources. Taking into consideration the cloud computing bank with its somehow commercial nature, the resources unavailability, such as liquidity risk, remains. In this paper, an attempt to show through a solution so far applied in economy, how would it be possible to predict such a liquidity risk in cloud computing bank. The proposed solution can especially be adapted to stock management. To reduce the risk we will also make use of a method inspired from physics based on the fluids mechanics; it is an application of Bernoulli's theorem called Torricelli. The resource bank will be considered as a reservoir of liquid, and the availability of resources then will depend on the liquid flow velocity and the replacement.

## 1. Introduction

The cloud computing is a real revolution; it represents a fundamental shift in the way services are delivered. Unlike the grid computing, cloud computing services can be adapted to the real human and social needs and easily commercialized. It is an economic model which provides a pool of resources, accessible for anyone, at any time and from everywhere. The provider, in cloud computing, is responsible for supplying the resources (infrastructure, servers, storage, and network), and the end-users take advantage of cloud services by paying for the used resources [10].

All firms may have to cope with risks and since cloud computing is a resource bank, like any commercial bank, it can at any time face risk, particularly liquidity risk or simply unavailability of services [1]. Managing such a risk is unavoidable that is proper analysis and appropriate action must follow.

The high occurrence probability of liquidity risk forces the commercial bank to be in a complete state of readiness to overcome such a situation. One can define the liquidity as the capacity, for a given entity, to face its obligations in due time, and thus enabling the treasury capacity to support the negative money flows within its duties towards its clients [1]. On the other hand, the liquidity risk is apparent when the bank does not have the necessary funds to honor all its engagements in due time.

In order to preserve the bank viability, each bank, as Basel committee explains [13], ought to dispose a rigorous process to identify measure, monitor and control the liquidity risk. It should establish a good funding strategy, ensuring effective diversification of sources and financing forms, taking into consideration that insufficient liquidity can rapidly exhaust a bank while the liquidity excess may slowly kill it [3]. Periodically, the Bank has to make crisis simulations on various stress scenarios, brief or prolonged, to test and adapt its strategies to liquidity risk management. Contingency funding plan is required to resolve liquidity shortages in the emergency event.

According to Harrington [2], a bank liquidity risk can be found at three levels:
- **Funding risk:** A risk that is manifested by the necessity of having new resources when the bank resources are no more available particularly after the massive withdrawals.
- **Time risk:** A risk which may appear when the bank doesn't succeed in making the expected money receipts (client's inability to repay the loan).
- **Call risk:** A risk which is related to the acquisition of new resources as a result of important loans in credit line.

Having said that it becomes necessary for all banks to establish a robust framework so that it can ensure a sufficient liquidity at all times and face this variety of risks.

In this paper we suggest the use of a method inspired from physics to palliate the problem of non availability of resources and determine an acceptable liquidity level for our bank. The method is based on the fluids mechanics. We consider the resource bank as a reservoir of liquid, and the availability of resources will then depend on the liquid flow velocity.

We start this paper by a short introduction presenting the new economic model and their risks. Our suggestion is presented and detailed in the second section, and finally we present the experimental study with all necessary details.

## 2. Fluids mechanics

Fluids mechanics is a science concerned with the laws of fluids flow. It's a branch department in physic which studies the fluid flows when these latter face forces and constraints. According to Hamouda [6], the fluids mechanics can be decomposed into two big branches, the static fluids which studies the fluid in rest and dynamic fluids which studies the fluid in movement [6].

A fluid can be considered to be formed of a great number of particles, very small and move freely among themselves. It characterized by: voluminal weight, the density and viscosity [8]. It is a continuous physical environment which can be deformed, not rigid and which can flow out [9].

The fluids may classified into:
- **Perfect fluids** when it is possible to describe their movement without considering the rubbing effects,
- **Real fluids** (without rubbing),
- **Incompressible fluids** (irreducible), when the voluminal mass does not depend on the pressure or the temperature (liquid),
- **Compressible fluids** (gas) [6].

In this article, we try to discuss the Bernoulli's theorem whose origin within the energy conservation with relation to the perfect fluid in a rest state. In 1937 it was established by Daniel Bernoulli through expressing the fluid's simplified hydraulic assessment in a conduit. It helps determining the relation between pressure, density and velocity at any point within a fluid, which can be calculated by the use of following equation:

$$\rho \frac{V_1^2}{2} + \rho g z_1 + p_1 = \rho \frac{V_2^2}{2} + \rho g z_2 + p_2 \quad (1)$$

This theorem is excellent and helpful since it is a purely algebraic relation that connects the speed, pressure and position of the fluid [7]. The Bernoulli's theorem states that the flow-wither perfect, incompressible, permanent (the parameters which characterize its «pressure, temperature, velocity, voluminal mass" have during time, a constant value), and the energy dissipation are negligible [7].

Torricelli's theorem is an application of Bernoulli's theorem which helps to define the empting velocity of a container with a certain liquid height.

## 2. The flow of the liquid contained in a reservoir-Torricelli's theorem

Let's consider a reservoir with a small opening at its base, a section S and a current line going from the surface at point '1' arriving at the opening point '2'. The Torricelli' formula can easily be established through Bernoulli's equation (1) to calculate the reservoir empty velocity. If the orifice diameter is smaller than the reservoir diameter, then the flow velocity at the point '1' is negligible, the emptying is then rapid (v1<<v2, we then consider v1 ≈ 0) [7]. The flow velocity in the same as the free fall between the free surface and the orifice, whatever the liquid voluminal mass is. The following figure illustrates Torricelli's principle.

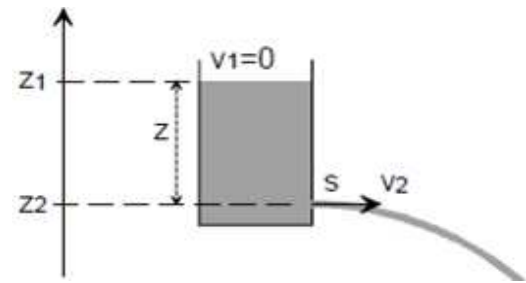

Fig. 1. A reservoir with a small opening at its base (parabolic jet).

In our study, we attempt to maintain the reservoir at a certain level and not allowing it empty in any case, for this we propose a method inspired from economy which gives the possibility of management of the liquid quantity that gets out of the reservoir. When the volume is under a certain level, we fill in it again. This, in finance, is called the stock replacement [1].

## 4. Stock management

A stock can be defined as a reserve intended to satisfy a subsequent need [4]. Optimizing a stock, means that it should be available at time without excess or insufficiency, to be able to reach the ideal compromise between minimum stocking cost and a maximum rate service.

A good stock management depends on the nature of replacement. In order to have the stock continuously operational (without falling short), we can:

- a- Either put in orders in precise date with always the same time limit between two orders: in this case we have "replacement within regular period",
- b- Or put in order when the stock drops to a certain level,
- c- Or even periodically defining a replacement program for a coming period taking into consideration the actual situation and all the estimates of prevision [5].

Replacing in the right moment is often difficult. The consumption before replacing [5] is either:

- a- Exactly as we have expected and the stock is nil at the moment of order,
- b- Or higher than expected and then we will be having a stock interruption before the time of order coming,
- c- Or a little but less than expected and we still have money in stock at the moment of making order.

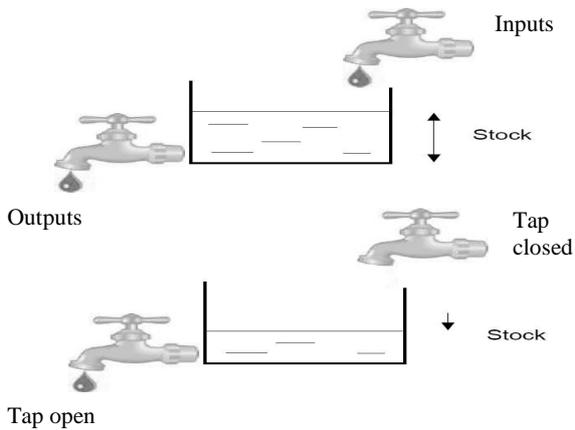

Fig. 2. Figure shows the stock when it gets ruptured [5].

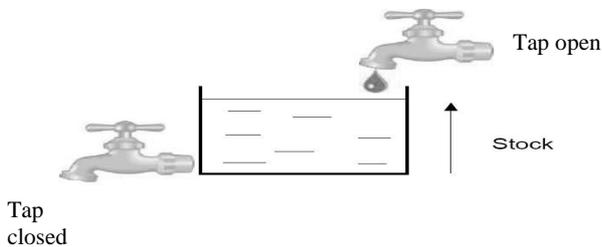

Fig. 3. Figure shows the stock when it is increasing and becoming invasive [5].

The security stocks usually overcome the stock interruption problem and ensure the continuity of activities. By adding to the security stock the expected consumption during the replacement time, it is possible to reach command point.

The command point may be equal to security stock plus expected quantity to be consumed during time delivery.

Figure 4 shows the replacement principle:

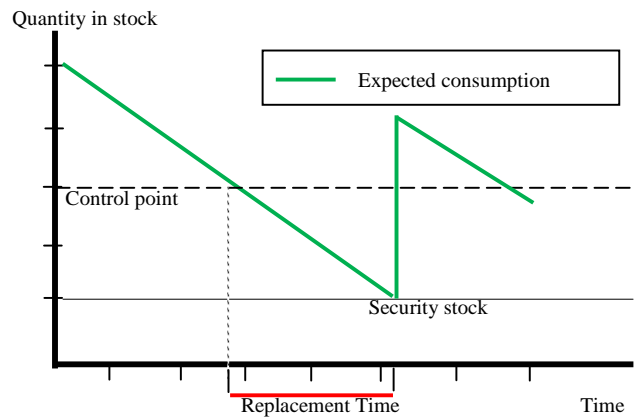

Fig. 4. The replacement.

To calculate the security stock, we take into account the statistical importance of demand variations, known through computing the standard deviation or the absolute average deviation (ASM) [5], and variations delivery.

According to Tchokogué [12], the choice of a service level defines the breaking risk which the manager can tolerate, it also help defining a level of security stock starting from "factor 'z'", given by the normal law table. Here some 'z' factors associated to rate service:

Table 1: Normal law table.

| Service rate | Risk of rupture | Factor 'z' |
|---|---|---|
| 50 | 50 | 0 |
| 84 | 16 | 1 |
| 90 | 10 | 1.3 |
| 95 | 5 | 1.6 |
| 97.5 | 2.5 | 2 |
| 99 | 1 | 2.3 |
| 99.5 | 0.5 | 2.6 |
| 100 | 0 | 3 |

By supposing that the demand is within a normal law, we can have three possible combinations to compute the security stock:

**Demand variation (only):** Security Stock (SS) therefore is:

$$SS = z \times \sigma_{dmdl}$$

With
$z$: is the security factor (number of standard deviation),
$\sigma_{dmdl}$: is the demand standard deviation during the replacement time.

**Time delivery variation (only):** demand variation during time becomes:

$$\sigma_{dmdl} = \sigma_{dl} \times \overline{dm}.$$

Where

$\sigma_{dl}$: is the time standard deviation.
$\overline{dm}$: average demand per unit of time.

Security stock SS become

$$SS = z \times \sigma_{dmdl}$$
$$SS = z \times \sigma_{dl} \times \overline{dm}$$

**Demand and time delivery variation (both):**

$$\sigma_{dmdl} = \sqrt{\sigma_{dm}^2 \times m + \overline{dm}^2 \times \sigma_{dl}^2}$$
$$SS = z \times \sqrt{\sigma_{dm}^2 \times m + \overline{dm}^2 \times \sigma_{dl}^2}$$

Where

$\sigma_{dmdl}^2$: required variation during the replacement cycle.
$\sigma_{dm}^2$: required variance.
$\sigma_{dl}^2$: period variance.

$$m = \frac{\text{length of time}}{\text{length of demand calculation period}}$$

The following figure shows the statistical properties of normal distribution curve,

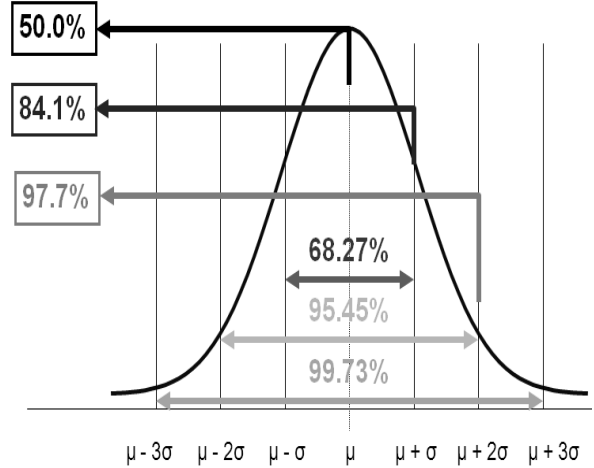

Fig. 5. Statistics of the normal distribution with μ variations in demand or time properties.

## 5. Experimental study

In order to evaluate the performance of our proposed algorithm, we discuss the results obtained by a simulation intended to objectively evaluate our contribution.

Based on the idea of finding a solution to liquidity problem and especially the cloud bank liquidity risk, we opted for the Torricelli's theorem and its applications as a means to efficiently manage the cloud bank data. This theorem helps supervising the way of emptying intelligently a reservoir.

The regularization of liquid quantity which gets out of the reservoir can either be done by adjusting the flow velocity or by adjusting of the orifice (s section). This adjustment act will naturally be based on the knowledge of the resources quantity in the reservoir.

To predict the resources quantity in the reservoir, we will make use of several stock management rules [1].

In this experimental study, we will engage some variables to test algorithm efficiency:

**Torricelli's parameters:**
- Bandwidth which plays the orifice role: if we wider the band then the size will increase.
- The flow velocity which defines the number of issued bits per second.

**Parameters of stock management rules:**
- The average demand per day and service level.
- The command point and the security stock depend on the variation demand or on the replacement time with claimed service level.

Since we've been considering the bandwidth as an opening of reservoir, the size of this latter is stated at the start, and the flow velocity is stable. The average demand per day is obviously known.

When the command point is reached we pass another command. If the command arrives at the right moment (i.e. before starting the security stock) no change would be expected. In case of any delay the bandwidth and the velocity should be adapted so that the security stock may resist till the command arrival.

If we suppose that our firm has enough stocking space, these experiences and results may follow:

**1. Demand variation**

Table 2: Command point by changing Service level.

| Demand per day | Service Level | Demand deviation | Security Stock SS | Command point PC |
|---|---|---|---|---|
| 100 GO | 99% (z=2.9) | 20 GO | 58 GO (SS=2.9 x 20) | 158 GO (PC=100+58) |
| 100 GO | 97.5% (z=2) | 20 GO | 40 GO (SS=2 x 20) | 140 GO (PC=100+40) |
| 100 GO | 95% (z=1.64) | 20 GO | 32.8 GO | 132.8 GO |

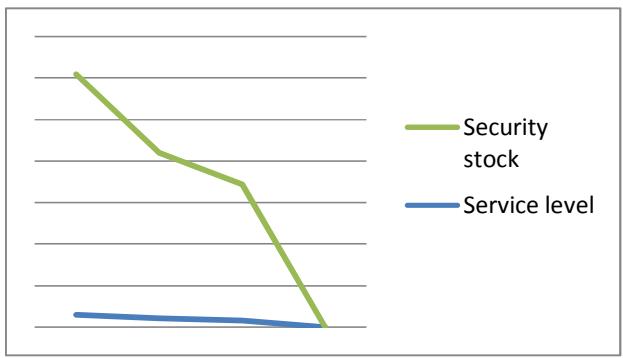

Fig. 6. security stock variation according to service level.

If the service level decreases, the security stock and command point decrease as well.

Table 3: Command point by changing Demand deviation.

| Demand per day | Service Level | Demand deviation | Security Stock SS | Command point PC |
|---|---|---|---|---|
| 100 GO | 97.5% | 30 GO | 60 GO | 160 GO |
| 100 GO | 97.5% | 20 GO | 40 GO | 140 GO |
| 100 GO | 97.5% | 15 GO | 30 GO | 130 GO |

The breaking risk for a service level of 97.5% equals 2.5. Figure 4 shows the normal distribution curve with a demand variation μ = 100 GO. With a security stock equal 40 GB, we can avoid a rupture and resist until the arrival of command.

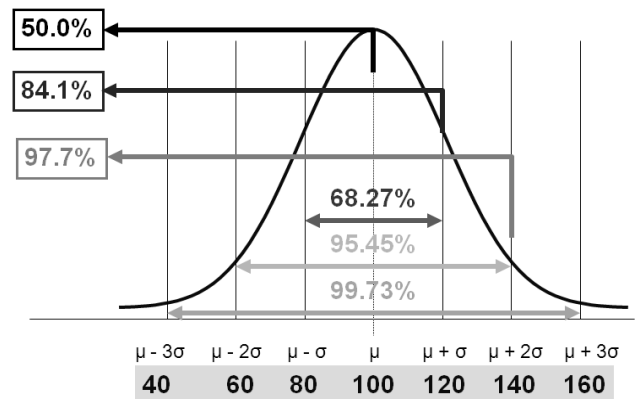

Fig. 7. Normal distribution curve with μ variation.

**2. Time variation (average replacement time is one day)**

Table 4: Command point by changing Service level.

| Demand per day | Service Level | Time deviation | Security Stock SS | Command point PC |
|---|---|---|---|---|
| 100 GO | 99% | 0.2 days | 58 GO SS=(2.9x0.2x100) | 158 GO PC=(100x1)+58 |
| 100 GO | 97.5 % | 0.2 days | 40 GO | 140 GO |
| 100 GO | 95 % | 0.2 days | 32.8 GO | 132.8 GO |

Table 5: Command point by changing Time deviation.

| Demand per day | Service Level | Time deviation | Security Stock SS | Command point PC |
|---|---|---|---|---|
| 100 GO | 97.5% | 0.8 days | 160 GO (SS=2x0.8x100) | 260 GO PC=(100x1)+160 |
| 100 GO | 97.5 % | 0.5 days | 100 GO | 200 GO |
| 100 GO | 97.5% | 0.2 days | 40 GO | 140 GO |

### 3. Demand and Time variation

Table 6: Command point by changing Time and demand deviation.

| Demand per day | Service Level | Demand deviation | Time deviation | Security Stock SS | Command point PC |
|---|---|---|---|---|---|
| 1024 GO | 98.75% (z=2.2) | 50 GO | 0.3 days | 684.7 GO | 1708.7 GO |
| 1024 GO | 98.75% | 30 GO | 0.2 days | 455.3 GO | 1479.3 GO |
| 1024 GO | 98.75% | 10 GO | 0.6 days | 1351.8 GO | 1451.8 GO |

For the previous tables the standard deviations are computed. The bandwidth and the flow velocity are stable. At this level the liquidity is being insured in advance by the security stock. If the consumption before the replacement is higher than expected and the stock quantity is equal to the security stock. The bandwidth and the velocity should be adapted that the security stock resist till the command arrival.

## 6. Conclusions

In this article, we propose two methods to insure the bank liquidity. The first method may be used in resources managing and in foreseeing liquidity risk. The second method based on fluids mechanics may be used to reduce the same risk. We consider the cloud bank as a reservoir of liquid which prevents it to become empty. The resources availability depends on the flow velocity and that of replacement. When the reservoir volume reaches a certain level we fill in it again. We can use Torricelli's theorem as a mean to insure the cloud bank liquidity and provide stock replacement method to predict liquidity risk.

### Acknowledgements

The author would like to thank Professor Mohamed Benyettou, for his very useful suggestions given during this work.

**First Author** Dou el kefel MANSOURI is a PHD student in department of computer science, University of Sciences and Technologies Mohamed Boudiaf USTO, ORAN-ALGERIA. His current researches are: cloud computing and scientific computing.

**Second Author** Mohamed Benyettou is a professor in department of computer science, University of Sciences and Technologies Mohamed Boudiaf USTO, ORAN-ALGERIA. He's also a director of laboratory of modeling and optimization LAMOSI, his main research interests include: artificial intelligence, optimization methods…